# Efficient Photocatalytic Hydrogen Production on Defective and Strained Black Bismuth (III) Oxide

Thanh Tam Nguyen[a,b] and Kaveh Edalati[a,b,]*

[a] WPI, International Institute for Carbon Neutral Energy Research (WPI-I2CNER), Kyushu University, Fukuoka 819-0395, Japan

[b] Mitsui Chemicals, Inc -.Carbon Neutral Research Center (MCI-CNRC), Kyushu University, Fukuoka 819-0395, Japan

Bismuth (III) oxide ($Bi_2O_3$) has been highly studied as a photocatalyst for green hydrogen production due to its low band gap, yet its efficiency requires enhancement. This study synthesizes a defective and strained black $Bi_2O_3$ by severe straining under high pressure, via a high-pressure torsion method, to improve its photocatalytic hydrogen production. The material rich in oxygen vacancies exhibits a ten-fold improvement in water splitting with excellent cycling stability. Such improvement is due to improved light absorption, narrowing band gap and reduced irradiative electron-hole recombination. Moreover, the valence band bottom energy positively increases by straining leading to a high overpotential for hydrogen production. This research highlights the potential of vacancies and lattice strain in developing dopant-free active catalysts for water splitting.

*Corresponding author (E-mail: kaveh.edalati@kyudai.jp; Tel: +80-92-802-6744)

# 1. Introduction

The production of renewable hydrogen is crucial for developing a clean energy system to deal with global warming and air pollution caused by carbon dioxide emission [1]. Currently, however, the primary method for hydrogen production is from fossil fuels, which are limited in supply and emit carbon dioxide, leading to severe global warming. Innovative sources for hydrogen production are necessary, with water being an ideal source [2]. Through various chemical reactions such as thermal, thermo-chemical, electrolysis and photochemical reactions, hydrogen and oxygen can be produced from water splitting. Among these methods, photocatalytic hydrogen production represents an ideal solution due to its simplicity, requiring only sunlight, water and catalyst, as well as its cleanliness and operation under mild conditions without emitting carbon dioxide [3]. Despite these environmental benefits, significant challenges must be addressed to make photocatalytic water splitting feasible. The primary challenge is developing efficient photocatalysts with appropriate electronic band structure, large light absorption, slow electron-hole recombination rates and effective electron separation and migration.

Since the first introduction of rutile for photocatalysis in 1972 [4], a variety of semiconductors, such as $TiO_2$ [5,6], $Cu_2O$ [7,8], $C_3N_4$ [9,10], ZnO [11,12], ZnS [13,14] and TaON [15] have been used for photocatalytic hydrogen production. Notable p-type semiconductor bismuth (III) oxide ($Bi_2O_3$) has six main crystallographic polymorphs denoted by α, β, γ, δ, ε and ω corresponding to the monoclinic, tetragonal, bcc, fcc, orthorhombic and triclinic crystalline forms, respectively (bcc: body-centered cubic, fcc: face-centered cubic) [16,17]. $Bi_2O_3$ has a low band gap, good stability and great conductivity, and is toxic-element-free, which makes it a popular choice for photocatalytic processes. Among these polymorphs, α and β forms have narrower band gaps, resulting in broad light absorption and high photocatalytic performance. However, the β polymorph is metastable, while the α polymorph is stable at room temperature, making α-$Bi_2O_3$ a promising photocatalyst for water splitting [18,19]. Despite its potential, the photocatalytic performance of pure $Bi_2O_3$ is rather low due to low separation efficiency and rapid recombination of charge carriers.

Various techniques have been used to enhance the properties and activities of photocatalysts, such as introducing heterojunctions [20], developing mesoporous catalysts [21], doping [22], strain engineering [23,24] and defect engineering [25]. These methods aim to improve light absorption, facilitate separation of electrons and holes, minimize recombination of electrons and holes and optimize electronic band structure, thereby boosting photocatalytic activity. Defect and strain engineering as dopant-free approaches, in particular, significantly influence these photocatalytic features. Severe plastic deformation methods like equal-channel angular pressing/extrusion (ECAP) [26], high-pressure torsion (HPT) [27,28] and accumulative roll-bonding (ARB) [29,30] are prominent defect/strain engineering techniques. These methods introduce strain and various defects, including dislocations, vacancies, twins and grain boundaries, facilitate phase transformations and control the texture, ultimately enhancing mechanical and functional properties [31-33]. HPT, in particular, is effective for introducing defects and strain in oxides due to its high processing pressure [33]. In HPT (Figure 1a), powders are compacted in the form of a disc and processed between two massive anvils under several giga-Pascal pressures by rotation of anvils against each other for $N$ turns (larger $N$ introduces larger external strain and thereby larger defects into the material) [27,28]. HPT-processed oxides have already been used in electrocatalytic [33], photovoltaic [34] and photocatalytic applications [35].

This study employs the HPT method for defect and strain engineering of α-$Bi_2O_3$ to enhance photocatalytic hydrogen production. The resulting catalyst demonstrates significant light absorption, a narrow band gap, reduced electron-hole recombination and the presence of oxygen vacancies. Consequently, the defective and strained catalyst exhibits more than ten-fold improvement in hydrogen production performance compared to the initial α-$Bi_2O_3$ powder.

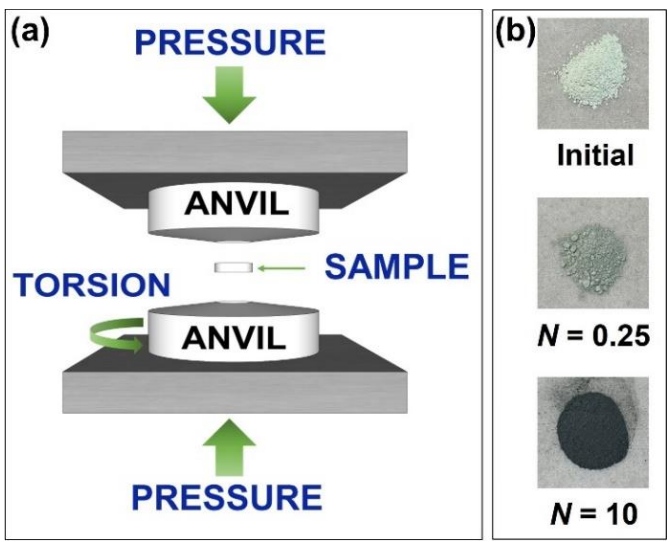

Figure 1. (a) Illustration of severe plastic deformation via HPT and (b) appearance of initial $Bi_2O_3$ powder and HPT-treated samples after $N = 0.25$ and $N = 10$ turns.

## 2. Experiment

### 2.1. Catalyst Synthesis

The initial $Bi_2O_3$ powder with 99.9% purity was purchased from Kojundo, Japan. To synthesize defective $Bi_2O_3$ by HPT, 250 mg of $Bi_2O_3$ powder was compressed under 380 MPa into a 10 mm diameter disc. The disc was then placed in a 10 mm diameter and 0.25 mm deep hole in the lower anvil of an HPT machine. The anvil located at the bottom was raised to touch the anvil located on the top, applying a pressure of 6 GPa to the sample. The HPT process involved rotating the lower anvil at a speed of 1 rpm under 6 GPa pressure for $N = 0.25$ (low applied external strain) and $N = 10$ (high applied external strain) turns. After the HPT treatment, the samples were crushed into powders and subjected to various characterization and photocatalytic measurements.

### 2.2. Catalyst Characterization

The crystalline structure of the three catalysts was analyzed using X-ray diffraction (XRD) with Cu K$\alpha$ irradiation and a wavelength of $\lambda = 1.542$ Å. Raman spectroscopy, employing a laser source with $\lambda = 352$ nm, was used for investigating crystalline structures. The morphology of samples was characterized by scanning electron microscopy (SEM). The particle size distribution was measured by a laser diffraction particle size analyzer. Light absorbance of the three catalysts was measured by ultraviolet-visible (UV-vis) diffuse reflectance spectroscopy within the range of 200-900 nm. Band gap was determined by the Kubelka-Munk analysis. X-ray photoelectron spectroscopy (XPS), using an Al K$\alpha$ source, was conducted to study oxygen vacancies and estimate the bottom of the valence band, with XPS positions corrected based on the C 1s peak at 284.8 eV. Photoluminescence (PL) emission spectroscopy at the steady state, using a laser with $\lambda = 352$ nm, was employed to examine radiative electron-hole recombination.

### 2.3. Photocatalytic Experiments

The photocatalytic experiments were conducted under a xenon lamp (300 W) with a measured power of 18 kW/m$^2$. In the photoreactor, 50 mg of catalyst was added to 27 cm$^3$ $H_2O$, 3 cm$^3$ $CH_3OH$ and 0.25 cm$^3$ $Pt(NH_3)_4(NO_3)_2$ (0.1M). The system was purged with argon gas to ensure the absence of oxygen in the reactor, which could interfere with the water-splitting reactions. The initial photocatalytic test was conducted for a single 180-minute cycle. To evaluate the reusability of the catalysts, five repeated tests were performed, with air evacuation and argon injection between each cycle. The production rates of hydrogen and oxygen were measured using gas chromatography employing a thermal conductivity detector (TCD-GC) and a molecular sieve column.

## 3. Results

### 3.1. Catalyst Characterization

Figure 1b illustrates the appearance of three samples: initial $Bi_2O_3$ (no external strain application) and HPT-processed $Bi_2O_3$ after $N = 0.25$ (low external strain application) and $N = 10$ (high external strain application). The initial powder is light yellowish white, but it darkens after HPT for 0.25 turns turning to black after 10 turns. The color change suggests the development of color centers like oxygen vacancies in the material [35,36]. Similar behavior has been reported in black $TiO_2$ [37], $ZrO_2$ [36], $Nb_2O_5$ [38] and $SrTiO_3$ [39].

Figure 2 shows the crystal structure of three samples using XRD and Raman techniques. As illustrated in Figure. 2a, the initial powder possesses a monoclinic phase of α-$Bi_2O_3$ with a minor amount of triclinic phase which might be formed during the production of powder or shortly after their storage under ambient atmosphere [35]. The monoclinic phase has a P21/c space group with lattice parameters of $a = 5.8496$ nm, $b = 8.1648$ nm, $c = 7.5101$ nm, $α = γ = 90°$ and $β = 112.977°$. After HPT, there is a noticeable peak broadening, which intensifies with the number of turns from $N = 0.25$ to $N = 10$. Grain boundaries and other planar imperfections are the causes of this broadening [40]. All Raman peaks in Figure 2b correspond to the monoclinic phase, which is consistent with XRD profiles, in which the fraction of triclinic as the second phase is minor. The higher magnification of the Raman spectra in Figure 2c shows the peak shift after HPT, indicating that lattice expansion occurs in the sample due to the lattice strain [41].

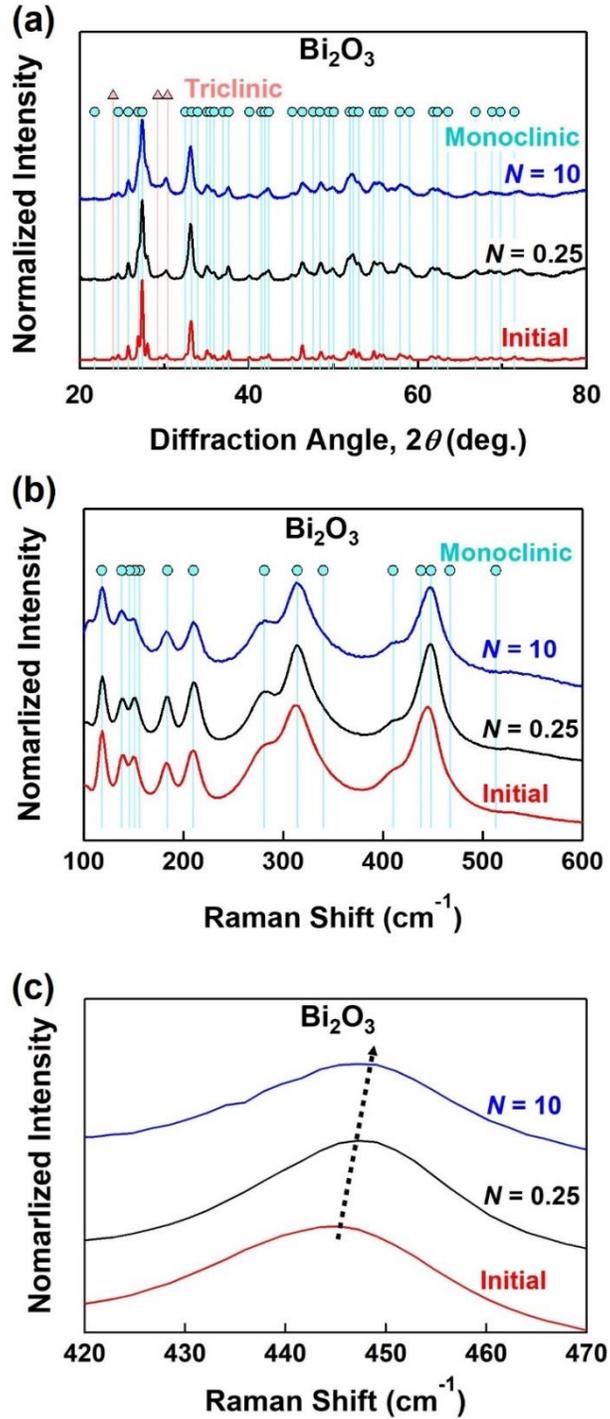

Figure 2. Stability of monoclinic phase after HPT processing. (a) XRD and (b,c) Raman analyses of initial $Bi_2O_3$ and HPT-treated samples processed for $N = 0.25$ and $N = 10$ turns, where (c) is a magnified view of (b) to show peak shifts.

The morphology of samples examined by SEM and the size distribution measured by laser diffraction particle size analyzer are shown in Figure 3. Both SEM and size distribution data are consistent, indicating the increase in particle size after HPT. The increase in particle size becomes more significant with the increase in HPT number of turns. The measured average particle sizes are 2.0, 3.7 and 21.7 µm for the initial powder and HPT-treated samples after $N = 0.25$ and $N = 10$, respectively. The surface area of samples was calculated by considering the mean particle size determined by the laser diffraction particle size analyzer as follows.

$$\text{Surface area in m}^2\text{ g}^{-1} = 6 / [\text{mean size of particles in µm} \times \text{density in g cm}^{-3}] \quad (1)$$

The calculated surface areas are 0.35 m$^2$ g$^{-1}$ for the initial Bi$_2$O$_3$ powder, 0.18 m$^2$ g$^{-1}$ for the HPT-treated sample after $N = 0.25$ and 0.03 m$^2$ g$^{-1}$ for the HPT-treated sample after $N = 10$. The increase in particle size results in a reduction in surface area, which is due to the powder consolidation by severe plastic deformation, also reported in various metals and ceramics earlier [42,43].

Figure 4a shows the UV-Vis spectra of the initial Bi$_2$O$_3$ and the samples treated with HPT for $N = 0.25$ and $N = 10$. The light absorbance increases after HPT processing, with the most significant increase is observed after $N = 10$. Figure 4b presents the Kubelka-Munk analysis used to estimate the direct band gap. A slight band gap narrowing occurs between the starting Bi$_2$O$_3$ and the sample treated with HPT under $N = 0.25$, which is around 2.8 eV. There is a further decrease in the band gap of the sample processed by HPT for $N = 10$ to 2.6 eV. The band gap narrowing is attributed to lattice defects and strain induced by severe plastic deformation [44]. The valence band top was determined using XPS analysis and shown in Figure 4c. By subtracting the band gap, calculated from the Kubelka-Munk plot, from the valence band top, the conduction band bottom was found. The electronic structures, including band gap and band edges (i.e. the top of the valence band and the bottom of the conduction band) of all samples, are shown in Figure 4d. The band structure of all samples meets the two energy conditions required for splitting water: (i) the valence band top is below 1.23 eV vs. NHE for H$_2$O/O$_2$ transformation, and (ii) the conduction band bottom is above 0 eV vs. NHE for H$^+$/H$_2$ transformation. The sample processed by HPT at $N = 10$ has the narrowest band gap, which is potentially the most beneficial for photocatalytic activity. Figure 4d also shows that in addition to band gap narrowing, the HPT process increases the energy level at the bottom of the conduction band, desirably resulting in a higher overpotential for hydrogen production.

To study the generation of oxygen vacancy-type defects, oxidation states and the recombination of photo-induced electrons and holes, XPS and PL analyses were conducted and presented in Figure 5. XPS spectra in Figure 5a indicate the oxidation state of $Bi^{3+}$, with a slight peak shift to lower binding energies in the HPT-processed samples, signifying the formation of oxygen vacancies. Furthermore, Figure 5b presents the oxygen XPS spectra and the oxidation state of $O^{2-}$. The initial powder shows one XPS peak which should correspond to lattice oxygen. For the HPT-processed samples, a shoulder appears at high energies and the intensity of this shoulder increases with increasing the number of HPT turns from 0.25 to 10. This shoulder should be due to the formation of oxygen vacancies [45,46]. In addition to XPS data, the change in color from yellow to black further suggests the formation of these HPT-induced oxygen vacancies. The steady-state PL spectra (Figure 5c) show the PL peaks at around 600 nm which correspond to the charge carrier radiative recombinations on defects in the samples. The intensity of PL in the initial $Bi_2O_3$ sample is higher than in the HPT-processed samples. The decrease in PL intensity becomes more significant with the increase in the number of HPT turns. This phenomenon, which was also reported in other HPT-processed oxides [47], indicates the suppression of electron-hole recombination by severe plastic deformation. Taken together, while the high light absorbance after HPT processing (Figure 4a) indicates easier charge carrier separation, the low PL intensity after HPT processing (Figure 4c) indicates less significant radiative recombination of charge carriers. Future studies using time-resolved PL can quantitatively clarify the effect of HPT on the migration and lifetime of charge carriers.

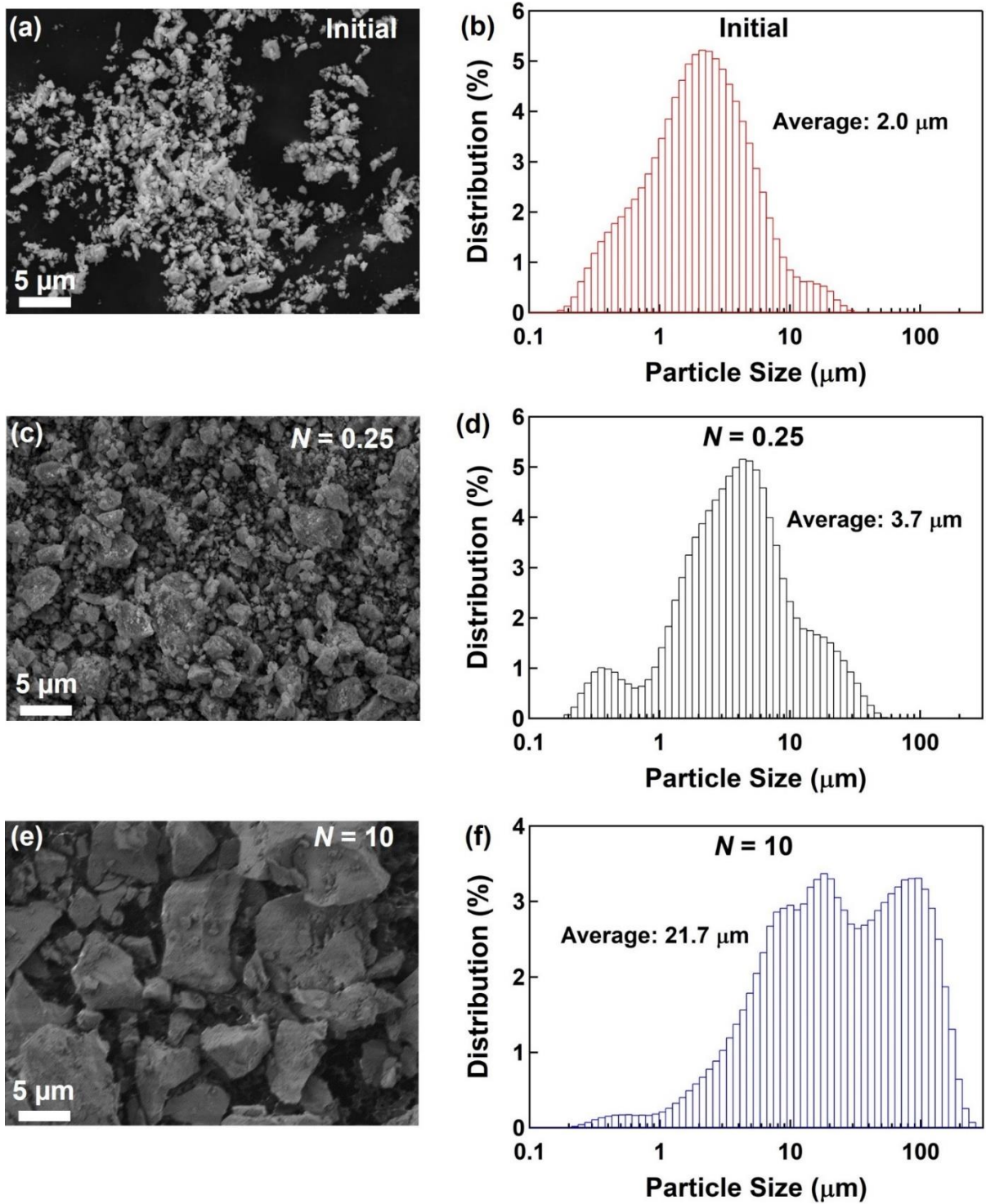

Figure 3. Increasing in particle size after HPT. SEM images and size distribution examined by laser diffraction particle size analyzer of (a, b) initial powder, (c, d) HPT-treated sample after $N = 0.25$ turns and (e, f) HPT-treated sample after $N = 10$ turns.

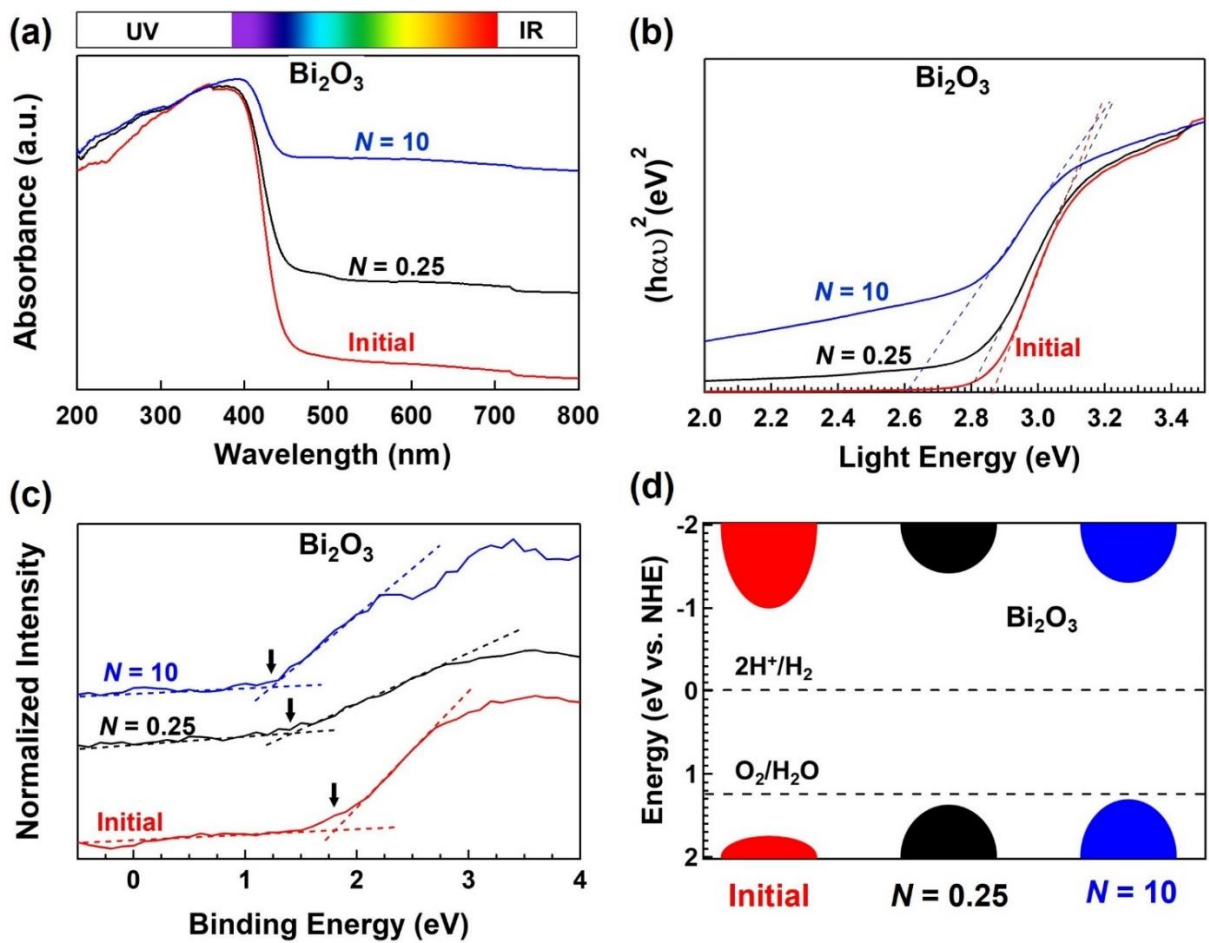

Figure 4. Large light absorbance and narrowing of optical band gap of $Bi_2O_3$ after HPT. (a) UV-vis absorption spectra, (b) band gap calculation by Kubelka-Munk method ($\alpha$: light absorbance, $h$: Planck's constant, $v$: frequency of light photons), (c) XPS of valence band top and (d) electronic structure of initial powder and HPT-treated samples after $N = 0.25$ and $N = 10$ turns.

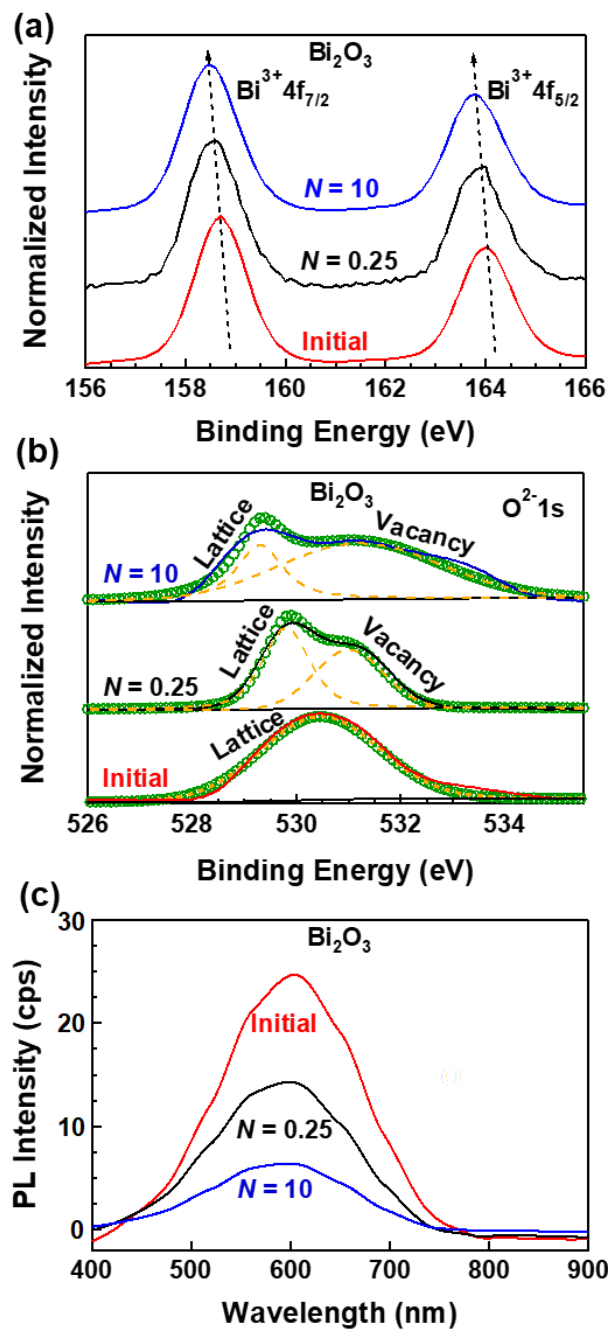

Figure 5. Formation of oxygen vacancies and reduction of electron-hole recombination by HPT processing in $Bi_2O_3$. (a) XPS spectra of bismuth element, (b) XPS spectra of oxygen element, and (c) PL spectra for initial powder and HPT-treated samples after $N = 0.25$ and $N = 10$ turns.

**3.2. Photocatalysis**

Photocatalytic hydrogen production experiments were conducted using the full arc of a xenon light, platinum as a co-catalyst and methanol as a sacrificial agent. Figure 6 illustrates the hydrogen production against irradiation time for various samples: a blank test (without catalyst), initial $Bi_2O_3$ powder and HPT-treated samples after $N = 0.25$ and $N = 10$ turns. Initially, no hydrogen is produced when the catalyst is stirred in the dark for one hour. Similarly, no hydrogen is detected in the blank test under light illumination without a catalyst. After 3 hours under light irradiation, 5.82 mmol m$^{-2}$ of hydrogen is produced in the system with the initial $Bi_2O_3$ powder. The hydrogen production slightly increases to 6.87 mmol m$^{-2}$ for the HPT-processed sample after $N = 0.25$ turns. As the number of turns increases to $N = 10$, the hydrogen production significantly increases more than 10 times to 59.6 mmol m$^{-2}$. This significant enhancement is noteworthy as it demonstrates the potential of a dopant-free strategy, which is currently a topic of considerable interest in the field [48,49].

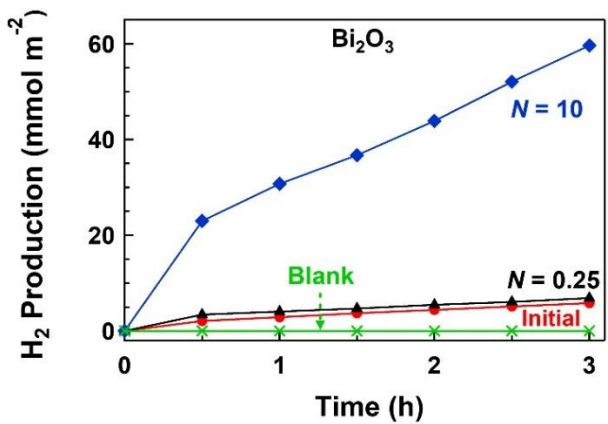

Figure 6. Enhancement of hydrogen production of $Bi_2O_3$ after HPT. Hydrogen production versus irradiation time during photocatalytic process for initial powder, HPT-treated samples after $N = 0.25$ and $N = 10$ turns, and blank test.

The post-catalysis sample after HPT processing for $N = 10$ turns was characterized by XRD as shown in Figure 7a. XRD spectra show a slight phase transformation from monoclinic to triclinic. One question raised from these phase transformations by irradiation: can the HPT-processed $Bi_2O_3$ catalyst work efficiently for a long time of photocatalytic hydrogen production? Therefore, five continuously repeated tests of photocatalysis were performed. As can be seen in Figure 7b, there is no reduction in hydrogen production after five repeated tests. Figure 7c exhibits an SEM image of the sample treated by HPT for $N = 10$ turns after five repeated photocatalytic cycles. The material after photocatalysis is similar to the one before photocatalysis in terms of size and shape. All data presented in Figure 7 indicate the excellent reusability of the defective and strained Bi$_2$O$_3$ catalyst for photocatalytic hydrogen production.

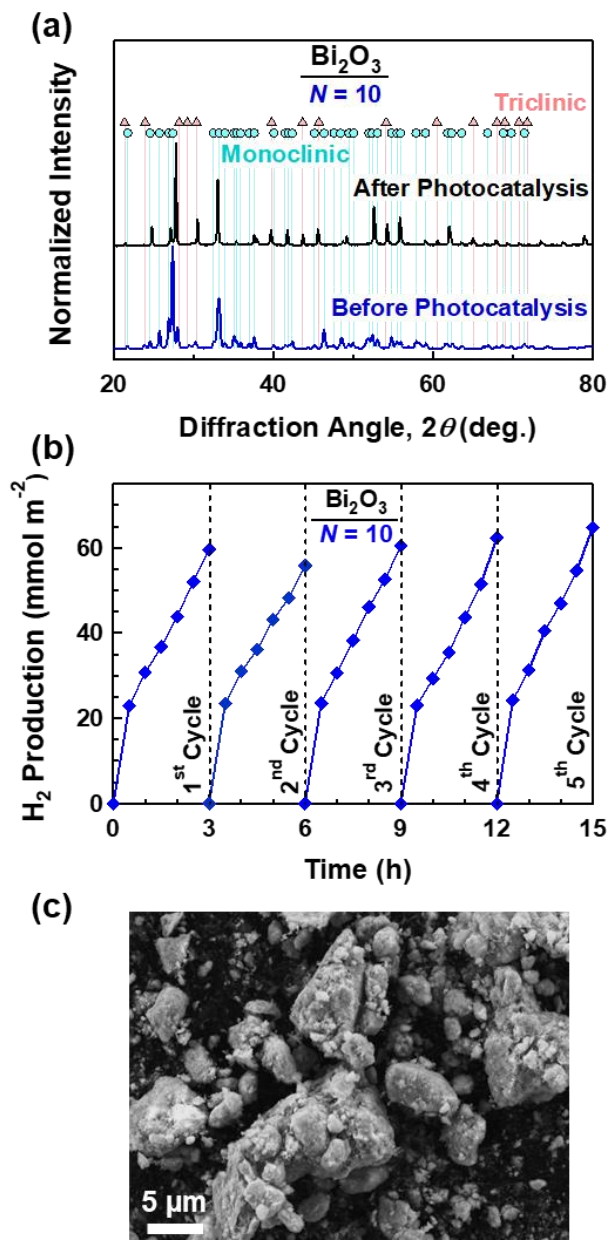

Figure 7. High reusability of HPT-processed Bi$_2$O$_3$ for photocatalytic hydrogen production. (a) XRD spectra of HPT-processed Bi$_2$O$_3$ after $N$ = 10 turns before and after photocatalysis, (b) hydrogen production versus irradiation time for five repeated cycles, and (c) SEM image of sample treated by HPT for $N$ = 10 turns after five repeated photocatalytic cycles.

## 4. Discussion

This study, in good agreement with earlier studies [18,19], shows that $Bi_2O_3$ can be successfully employed for photocatalytic hydrogen production. A defective and strained $Bi_2O_3$ synthesized by the HPT process has higher photocatalytic activity compared to the initial powder. The amount of hydrogen production is enhanced with the enhancement of HPT turn numbers. The reasons for such superior photocatalytic activity are large light absorbance, narrow band gap, high conduction band bottom energy (i.e. higher overpotential for hydrogen production) and reduced electron-hole recombination. The formation of lattice strain and oxygen vacancies, which is confirmed by the color change and Raman and XPS peak shifts, appears to be responsible for such modification.

The key advantage of this study lies in the engineering of oxygen vacancies in $Bi_2O_3$ photocatalysts without introducing impurity atoms. Oxygen vacancies offer significant benefits for photocatalysis, such as expanding the light absorption, adjusting the band structure, accelerating the migration and separation of electrons and improving reactions on the outer layer of the catalyst [45,46,50-52]. However, creating oxygen vacancy-type defects using impurities can undermine such advantages because of impurity-driven recombination. In general, the excess electrons generated from oxygen removal typically migrate to the empty orbitals of metals, leading to the generation of shallow donors below the conduction band bottom and increasing absorption of light photons [53]. Furthermore, oxygen vacancies on the surface can enhance charge carrier separation by serving as locations for photocatalytic reactions. Vacancy-type defects generated without impurity atoms are anticipated to be more efficient for generating active atomic-scale sites due to their diverse coordination numbers and dangling bonds [51,53].

Studies have indicated that both bulk and surface oxygen vacancies are essential for enhancing photocatalytic activities and reducing the impact of vacancies on the recombination of photo-induced electrons and holes [54]. The HPT method effectively introduces oxygen vacancies throughout the material, including both surface and bulk. The creation of these vacancies through HPT is attributed to the strain-induced generation of point defects and the pressure-induced suppression of effect annihilation [55]. In contrast, oxygen vacancies in chemically synthesized black oxides are typically formed only on free surfaces or in distorted phases close to free surfaces [56,57]. Therefore, the HPT method, a severe plastic deformation technique, is shown to be a superior dopant-free approach for the introduction of oxygen vacancies, thereby significantly enhancing photocatalytic activity [58]. Although photocatalytic test data in different publications should be compared with care due to the differences in reactors, light sources, additives and measurement systems, the amount of hydrogen production from $Bi_2O_3$ in this study is higher than many data reported for catalysts processed by HPT [12,35,36,59-62] or other methods [63-70], as compared in Table 1.

Table 1. Comparison of hydrogen production from HPT-processed $Bi_2O_3$ in comparison with other catalysts produced by HPT or other processes.

| Catalyst | Process | Produced $H_2$ (mmol m$^{-2}$ h$^{-1}$) | Ref. |
|---|---|---|---|
| $Bi_2O_3$ | HPT | 19.88 | This study |
| $Ga_6ZnON_6$ | HPT | 0.023 | [35] |
| $ZrO_2$-3%$Y_2O_3$ | HPT | 0.032 | [36] |
| $TiO_2$ | HPT | 1.43 | [59] |
| $TiO_2$-ZnO | HPT | 0.0003 | [60] |
| ZnO | HPT | 0.50 | [12] |
| $TiZrHfNbTaO_{11}$ | HPT | 0.073 | [61] |
| $TiZrHfNbTaO_6N_3$ | HPT | 0.028 | [62] |
| $TiO_2$ | Sol-gel | 0.179 | [63] |
| $TiO_2$ | Hydrothermal | 0.031-0.074 | [63] |
| P25 | Commercial | 0.025-0.710 | [64] |
| P25 | Commercial | 0.284 | [65] |
| P25 | Commercial | 0.040 | [66] |
| $TiO2$-$CuO_x$ | Impregnation | 0.087 | [67] |
| $TiO_2$-ZnO | Sol-gel | 0.030 | [68] |
| ZnO | Sol-gel | 0.002-0.052 | [69] |
| ZnO-CuO | Sol-gel | 0.129-0.260 | [69] |
| ZnO-C | Calcination | 0.023 | [70] |

## 5. Conclusions

This study demonstrated a straightforward mechanical dopant-free HPT method to synthesize a strained and defective black $Bi_2O_3$ photocatalyst with large light absorbance, narrow band gap, small electron-hole recombination and beneficial oxygen vacancies for enhancing photocatalytic hydrogen production. By varying the number of HPT turns, we observed a corresponding increase in photocatalytic hydrogen production, highlighting the positive impact of HPT on catalyst properties and activities. These findings suggest that HPT is an effective method for producing highly active black catalysts for water-spitting applications.

**CRediT authorship contribution statement**

T.T. Nguyen and K. Edalati: Conceptualization, Methodology, Experimental, Writing – reviewing & Editing.

**Declaration of Competing Interest**

The authors declare that they have no known competing financial interests or personal relationships that could have appeared to influence the work reported in this paper.

**Acknowledgments**

This work is funded partly through a Grant-in-Aid from the Japan Society for the Promotion of Science (JSPS) (JP22K18737) and in part by the ASPIRE project of the Japan Science and Technology Agency (JST) (JPMJAP2332).This work is funded partly through a Grant-in-Aid from the Japan Society for the Promotion of Science (JSPS) (JP22K18737) and in part by the ASPIRE project of the Japan Science and Technology Agency (JST) (JPMJAP2332).